\documentclass[twocolumn,showpacs,superscriptaddress,preprintnumbers,amsmath,amssymb,epsfig,floatfix,prl]{revtex4-1}

\usepackage{amssymb}
\usepackage{epsfig}
\usepackage{graphicx}
\usepackage{dcolumn}
\usepackage{color}
\usepackage{natbib}  
\usepackage{hyperref}
\usepackage{breakurl}
\hypersetup{colorlinks=true, citecolor=blue, urlcolor=blue, linkcolor=blue}
\usepackage{bm}
\usepackage{tabularx}
\newcolumntype{L}[1]{>{\raggedright\arraybackslash}p{#1}}
\newcolumntype{C}[1]{>{\centering\arraybackslash}p{#1}}
\newcolumntype{R}[1]{>{\raggedleft\arraybackslash}p{#1}}
\begin{document}
\title{Giant Enhancement of Superconductivity in Zr Point Contacts}

\author{Mohammad Aslam}
\affiliation{Department of Physical Sciences, Indian Institute of Science Education and Research, Mohali 140306, India}
\author{Chandan K. Singh}
\affiliation{Department of Physics, Indian Institute of Science Education and Research, Pune 411008, India}

\author{Shekhar Das} 
\affiliation{Department of Physical Sciences, Indian Institute of Science Education and Research, Mohali 140306, India}
\author{Ritesh Kumar}
\affiliation{Department of Physical Sciences, Indian Institute of Science Education and Research, Mohali 140306, India}
\author{Soumya Datta} 
\affiliation{Department of Physical Sciences, Indian Institute of Science Education and Research, Mohali 140306, India}
\author{Soumyadip Halder} 
\affiliation{Department of Physical Sciences, Indian Institute of Science Education and Research, Mohali 140306, India}

\author{Sirshendu Gayen}
\affiliation{Department of Physical Sciences, Indian Institute of Science Education and Research, Mohali 140306, India}
\author{Mukul Kabir}
\email{mukul.kabir@iiserpune.ac.in} 
\affiliation{Department of Physics, Indian Institute of Science Education and Research, Pune 411008, India}
\affiliation{Centre for Energy Science, Indian Institute of Science Education and Research, Pune 411008, India}
\author{Goutam Sheet}
\email{goutam@iisermohali.ac.in} 
\affiliation{Department of Physical Sciences, Indian Institute of Science Education and Research, Mohali 140306, India}
\date{\today}

\begin{abstract}
\textbf{For certain complex superconducting systems, the superconducting properties get enhanced under mesoscopic point contacts made of elemental non-superconducting metals. However, understanding of the mechanism through which such contact induced local enhancement of superconductivity happens has been limited due to the complex nature of such compounds.  In this paper we present giant enhancement of superconducting transition temperature (T$_c$) and superconducting energy gap ($\Delta$) in a simple elemental superconductor Zr. While bulk Zr shows a critical temperature around 0.6\,K, superconductivity survives at Ag/Zr and Pt/Zr point contacts up to 3\,K with a corresponding five-fold enhancement of $\Delta$. From first principles calculations we show that the enhancement in superconducting properties can be attributed to a modification in the electron-phonon coupling accompanied by an enhancement of the density of states which involves the appearance of a new electron band at the Ag/Zr interfaces.    }

\end{abstract}

\maketitle

Superconductivity emerging at the interfaces between two non-superconducting materials\cite{Go, Wu, Ko, Sm} and surprising enhancement of superconducting properties at mesoscopic interfaces has attracted significant attention recently. The  examples include (but, not limited to) (a) SrTiO$_3$/LaAlO$_3$\cite{Re}, where an interfacial 2D electron gas superconducts at sub-kelvin temperatures, (b) point contacts on FeSe and Sr$_2$RuO$_4$ , where a several-fold enhancement of $T_c$ is observed\cite{Yu, He, Ki, Pa, Gs, Bu, Yuk, Yin}, and (c) point contacts made on topological semimetals, where a robust superconducting phase is seen to emerge\cite{Lag, Agg}. In all these cases, achieving a clear understanding of the mechanism through which the enhancement/emergence of superconductivity happens remained elusive. The complex nature of these systems in terms of their band structure and the exotic electronic properties that they display makes the goal of such understanding non-trivial. However, observation of a similar effect in a simple elemental system is expected to facilitate such an understanding.

In this paper, we demonstrate an almost five-fold enhancement of the critical temperature in mesoscopic point contacts\cite{Nai, Lau, Gol} made on a very simple elemental superconductor Zr. Pure Zr shows a critical temperature around 0.57\,K \cite{zr1, Eis} in it's bulk form. Under point contacts on Zr made with pure elemental metals like Ag and Pt, the superconducting transition temperature dramatically increases to approximately 3\,K. This increase is accompanied by a corresponding increase in the superconducting energy gap. Observation of such a dramatic enhancement in superconducting properties at the mesoscopic interfaces made on a simple elemental superconductor is expected to provide useful information that might be useful in understanding the effect. We have attempted to understand the enhancement of $T_c$ in superconducting Zr point contacts by first principles calculations. Based on our calculations we conclude that in case of Zr, the enhancement can be attributed to two factors: (a) a significant enhancement of the DOS at the Fermi surface which is also associated with the emergence of a new electron-band at the Fermi surface, and (b) a substantial increase in the electron-phonon coupling\cite{Jess} under the point contacts, as compared to bulk Zr.

\begin{figure}[h!]
\centering
	\includegraphics[width=0.4\textwidth]{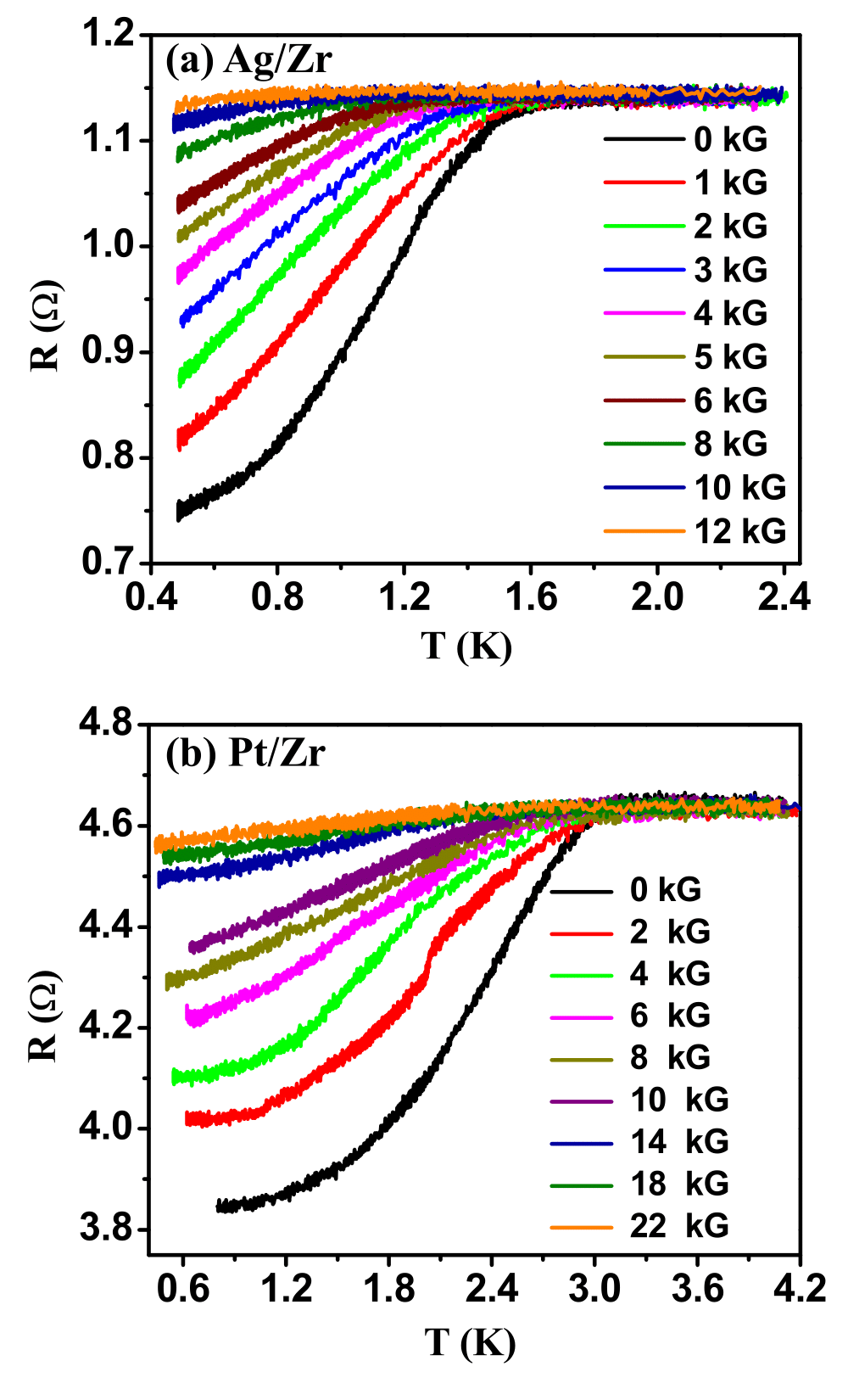}
	\caption{Resistance versus temperature ($R-T$) curves  for (a) Ag/Zr and (b) Pt/Zr point-contacts showing systematic drop of the superconducting transition temperature with varying applied magnetic field. }
	\label{f1}
\end{figure}

\begin{figure}[h!]
	\centering
	\includegraphics[width=0.5\textwidth]{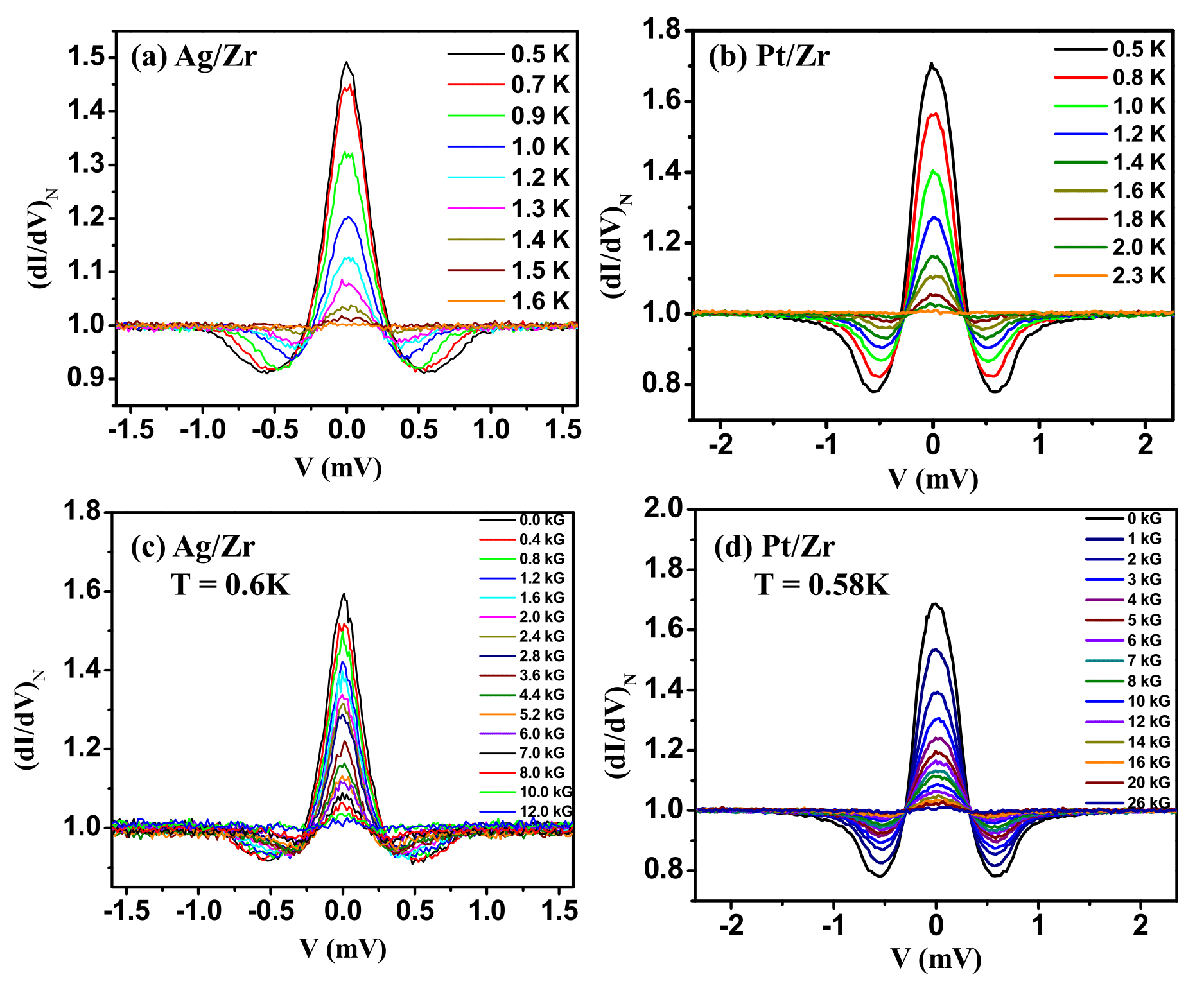}
	\caption{Temperature dependence of thermal limit spectra on (a) Ag/Zr and (b) Pt/Zr point-contacts. Magnetic field dependence of thermal limit spectra on (c) Ag/Zr and (d) Pt/Zr point-contacts.}
	\label{f2}
\end{figure}

The low-temperature experiments were performed in a He$^3$ cryostat working down to 500\,mK using a home-built point-contact spectroscopy probe. The probe is equipped with a piezo walker that facilitates fine adjustment of the tip-sample distance. The entire measurement system including the tip and the sample go to the center of a solenoidal magnet working up to 7\,Tesla. In Figure \ref{f2} we show point contact spectra obtained for (a,\,c) Ag/Zr and (b,\,d) Pt/Zr point contacts. At the lowest temperature (0.5\,K), the $dI/dV$ vs. $V$ spectra show a central peak followed by two dips symmetric about $V=0$. These spectral features indicate that the point contact is in the thermal regime of transport where the dips appear due to the critical current of the superconducting point contacts. All these features systematically evolve with increasing magnetic field and the features disappear at 1.6 K for Ag/Zr point contacts and at 2.3\,K for Pt/Zr point contacts. Temperature dependent measurements of the normal state resistance of these point contacts reveal superconducting transitions at 1.6\,K (for Ag/Zr) and 2.8\,K (for Pt/Zr) respectively (refer to Figure \ref{f1}). These values are significantly larger than the known bulk critical temperature of Zr (0.57\,K). The transition temperature goes down with increasing magnetic field, as expected for superconducting point contacts. We believe, for Pt/Zr point contacts, the spectral features disappear at a temperature lower than the resistive transition temperature due to associated thermal broadening of spectral features. The spectral features shown in Figure \ref{f2}\,(a,\,b) also fade away with increasing magnetic field.

\begin{figure}[h!]
\centering
	\includegraphics[width=0.5\textwidth]{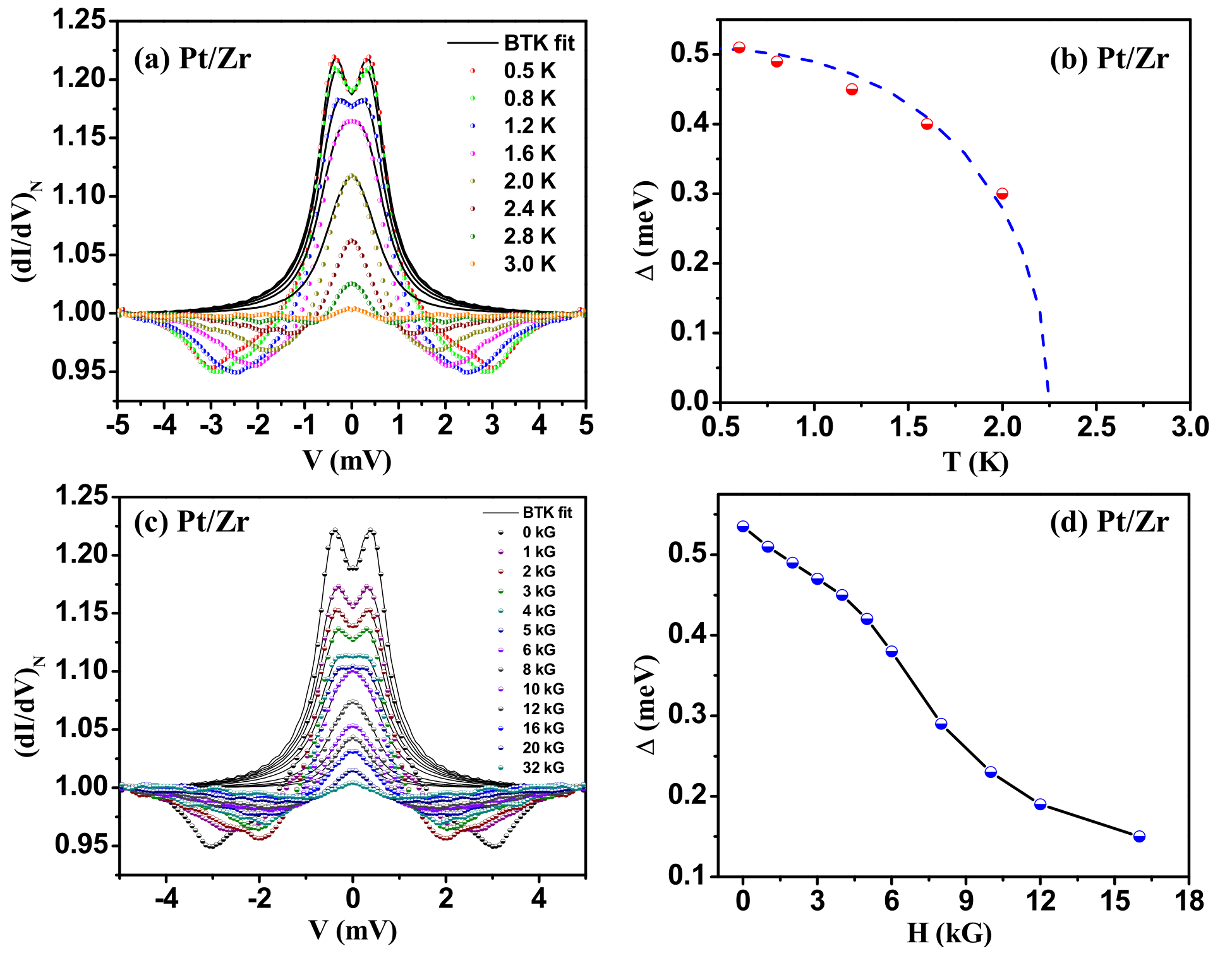}
	\caption{(a) Temperature dependence of the intermediate regime spectra on Pt/Zr point-contact. (b) Temperature dependence of the superconducting gap amplitude extracted from fitting the data following the BTK theory. (c) Magnetic field dependence of the intermediate regime spectra on Pt/Zr point-contact. (d) variation of the superconducting gap amplitude with magnetic field strength extracted from fitting the data following the BTK theory.}
	\label{f3}
\end{figure}

\begin{figure*}
\centering
\includegraphics[width=1\textwidth]{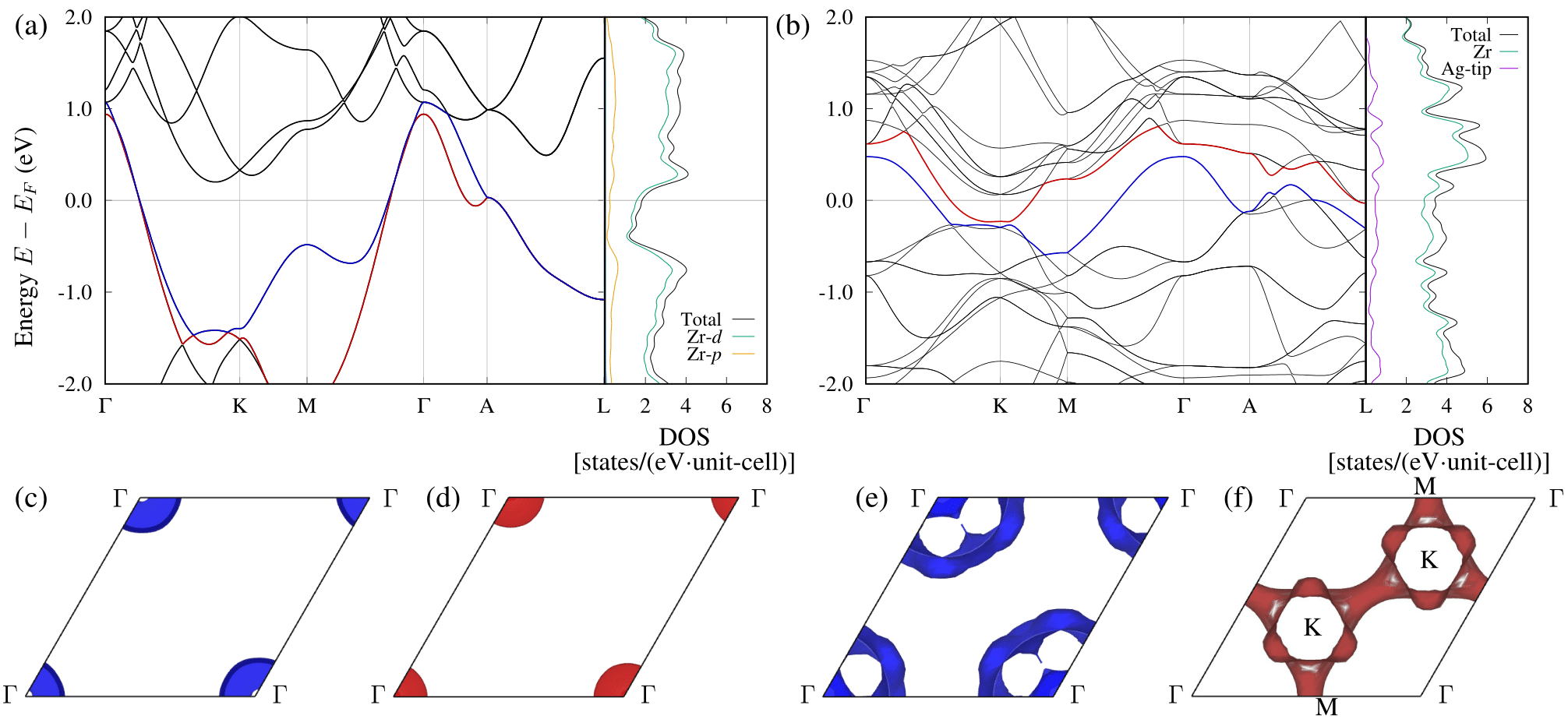}
\caption{Band structures and the corresponding density of states of (a) hcp-Zr and (b) hcp-Zr/Ag-tip. While two hole like bands crosses the Fermi level for hcp-Zr, an electron pocket appears due to the introduction of Ag-tip. The Zr-DOS at $E_F$ is increased by $\sim$62\% due to Ag tip insertion. The corresponding Fermi surfaces are shown in (c)-(f). The hole pocket in hcp-Zr/Ag-tip is larger in size, and the electron pocket at $K$-point are connected by a tubular network..  
}
\label{f4}
\end{figure*}

\begin{figure}[h!]

\centering
\includegraphics[width=0.5\textwidth]{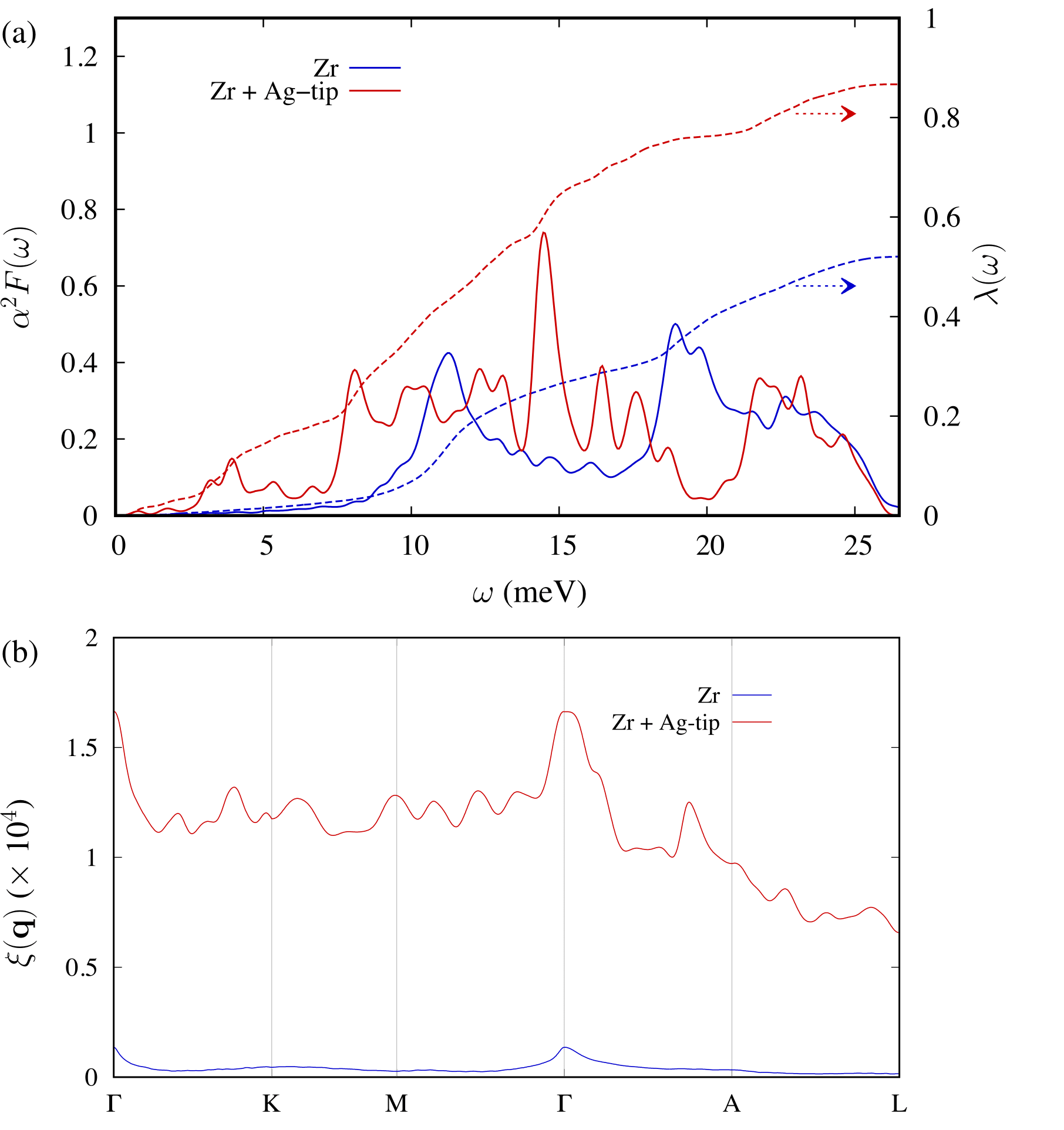}
\caption{(a) Calculated Eliashberg spectral function $\alpha^2F(\omega)$ and $\lambda(\omega)$. In addition to increased spectral weight, availability of many phonons to couple with electrons enhance $e-p$ coupling under Ag-tip. (b) An order of magnitude increase in the Fermi surface nesting function $\xi_{\mathbf q}$ for all $\mathbf q$-vectors inside the Brillouin zone contribute to increasing $\lambda(\omega \rightarrow \infty)$.
}
\label{f5}
\end{figure}

Though the critical temperature of the superconducting point contacts can be accurately estimated from the thermal regime data presented above, for the determination of superconducting energy gap ($\Delta$), the experiments must be performed in the ballistic or diffusive regime of transport where energy resolved spectroscopy is possible through Andreev reflection spectroscopy. We have driven the point contacts away from the thermal regime by reducing the diameter of the point contacts in-situ. As shown in Figure \ref{f3}\,(a) and (c), we have obtained data in a regime where clear features (two conductance peaks symmetric about $V=0$) associated with Andreev reflection appear. These data have also been fitted well by the model developed by Blonder, Tinkham and Klapwijk (BTK)\cite{Bl}. The BTK fits reveal a superconducting energy gap of 0.51\,meV for Pt/Zr point contacts indicating $\Delta/k_BT_c =$ 1.74, which is consistent with weak-coupling BCS prediction confirming that even above the bulk T$_c$ of Zr, the superconducting region confined under point contact remains conventional in nature. The Andreev reflection spectra evolve systematically with both temperature (Figure \ref{f3}\,(a)) and magnetic field (Figure \ref{f3}\,(b)). BTK analysis of the temperature dependent spectra confirm the BCS temperature dependence of the superconducting energy gap. The superconducting energy gap also decreases systematically with increasing magnetic field (Figure \ref{f3}\,(d)).
\\

To understand the microscopic origin of the giant enhancement of T$_c$ under mesoscopic point contacts on Zr, we have conducted first principles based calculations.  Electronic structure calculations were performed within norm-conserving Troullier-Martins pseudopotential scheme~\citep{PhysRevB.43.1993} and local-density approximation~\citep{PhysRevLett.45.566,PhysRevB.23.5048} implemented in QUANTUM ESPRESSO code.~\citep{0953-8984-21-39-395502} The Kohn-Sham wave functions and charge density were expanded in the plane wave basis with energy cutoffs of 60 and 240\,Ry. Structures were fully relaxed until the forces on each atom were smaller than 10$^{-4}$\,Ry/\AA\ using 12$\times$12$\times$8 Monkhorst-Pack $k$-mesh~\citep{PhysRevB.13.5188} with 0.02\,Ry Methfessel-Paxton smearing for electronic state occupancy.~\citep{PhysRevB.40.3616} Highly dense $k$-point grid of 60$\times$60$\times$60 was used to calculate Fermi surfaces.  Dynamical matrices are calculated on a 6$\times$6$\times$4 {\bf q}-mesh in the Brillouin zone using density functional perturbation theory (DFPT).~\citep{RevModPhys.73.515} Further, the electron-phonon ($e-p$) coupling was calculated using the electron-phonon Wannier method~\citep{PhysRevB.76.165108}  as implemented in the EPW code.~\citep{NOFFSINGER20102140} Maximally localized Wannier functions are constructed on a 6$\times$6$\times$4 $k$-mesh using Wannier90 code.~\citep{PhysRevB.56.12847,PhysRevB.65.035109,MOSTOFI2008685} Finally, fine electron 24$\times$24$\times$16 $k$-grid and phonon 24$\times$24$\times$16 $\bf q$-grids with phonon smearing 0.3\,meV were used  to calculate Eliashberg spectral function.

Phonon mediated superconductivity is well described by Migdal-Eliashberg theory and the corresponding $T_c$ is well approximated by the Allen-Dynes modified McMillan formula,~\citep{PhysRev.167.331,PhysRevB.12.905}
\begin{equation}
T_c = \frac{\omega_{\rm ln}}{1.2} \left[ \frac{-1.04(1+\lambda)}{\lambda -\mu^*(1+0.62\lambda)} \right],
\end{equation}
where $\omega_{\rm ln}$ is the logarithmic average of phonon frequencies, $\lambda$ is the $e-p$ coupling constant and $\mu^*$ is Coulomb pseudopotential.~\citep{PhysRev.125.1263} The Eliashberg spectral function for the $e-p$ interaction is obtained as,~\citep{PhysRevB.6.2577, PhysRevB.86.094515}
\begin{equation}
\alpha^2F(\omega) = \frac{1}{N(E_F)} \sum_{\mathbf{kq},\gamma} \left| \mathcal{G}_{\mathbf{k},\mathbf{k+q},\gamma}   \right|^2 \delta(\epsilon_{\mathbf{k}}) \delta(\epsilon_{\mathbf{k+q}}) \delta(\omega-\omega_{\mathbf{q},\gamma}), 
\end{equation}
where $N(E_F)$ is the electronic density of states (DOS) at the Fermi level, $\omega_{\mathbf{q},\gamma}$ is the phonon frequency of wave vector $\mathbf{q}$ and mode $\gamma$. The $|\mathcal{G}_{\mathbf{k},\mathbf{k+q},\gamma}|$ is the $e-p$ matrix element for the scattering of electron between states with momenta $\mathbf{k}$ and $\mathbf{k+q}$ at the Fermi level. Further $\lambda$ and $\omega_{\rm ln}$ are obtained using $\alpha^2F(\omega)$ as $\lambda = 2\int \frac{\alpha^2F(\omega)}{\omega} d\omega$ and $\omega_{\rm ln} = \exp[\frac{2}{\lambda} \int \frac{\alpha^2F(\omega)}{\omega} {\rm ln}(\omega) d\omega]$.
\\

\begin{table}[!t]
 \caption{Measured and calculated superconducting temperature using the Allen-Dynes modified McMillan formula, which is substantially enhanced under metallic point contact through increase in $e-p$ coupling. The theoretical $T_c$ is calculated using $\mu^*$=0.18.}
 \label{table1}
 \begin{tabular}{L{3cm}C{1.5cm}C{1.5cm}C{1.5cm}}
 \hline
 \hline \\[-1ex]
&  $\lambda$ & \multicolumn{2}{c}{$T_c$ (K)}  \\
&   &   Theory & Experiment \\
\hline\\ [-1.5ex]
hcp-Zr & 0.52 & 0.51 & 0.57 \\
hcp-Zr/Metallic tip & 0.87 & 3.52 & 3.2 \\ 
\hline
 \hline
 \end{tabular}
 \end{table}

We first discuss the picture for pure hcp-Zr, and subsequently investigate the effects of elemental Ag-tip. In Figure \ref{f4} (a), the decomposed band structure of hcp-Zr along the high-symmetry directions of Brillouin zone indicate two hole-like bands centered at the $\Gamma$-point, which originate mainly from the Zr-$d$ orbitals, while the Zr-p contribution at the Fermi level is negligible. These two bands become doubly degenerate at $A$-point. The corresponding Fermi surface (FS) consists of two cylindrical hole like sheets (Figure \ref{f4}(c), (d)) and the electronic DOS at $E_F$ is found to be 1.84\,states/[eV$\cdot$unit-cell] (Figure \ref{f4}(a)). The introduction of Ag-tip substantially perturbs the Zr band structure (Figure \ref{f4}(b)), which comprises of three hole pockets and one electron pocket. While the electrons around $K$-point and the hole pocket at $\Gamma$ are comparable in size, the hole pockets between $A-L$ are found to be much smaller. Further, the size of the hole pocket at $\Gamma$ is larger in size  (Figure \ref{f4}(e)) compared to pure-Zr, and the electron pockets at $K$ are connected by a tubular network (Figure \ref{f4}(f)). Due to the insertion of Ag-tip, the Zr-DOS at $E_F$ is increased by about 62\% to 2.98\,states/[eV$\cdot$unit-cell] (Figure \ref{f4}(b)). In contrast, a small contribution of 0.46\,states/[eV$\cdot$unit-cell] is contributed from the Ag-tip. Next, we compare the phonon dispersion and DOS without and with the tip (supplemental material). We do not observe any kinetic instability due to Ag-tip insertion, and interestingly observe softening of the acoustic modes and the additional appearance of many optical modes in the presence of tip.

In order to quantify the interaction between electrons and phonons, we calculate $\alpha^2F(\omega)$ and total $e-p$ coupling $\lambda(\omega \rightarrow \infty$), which are shown in Figure \ref{f5}(a).  In addition to increased spectral weight, more peaks in $\alpha^2F(\omega)$ between 5-25\,meV indicate more phonons to couple with electrons and enhance $e-p$ coupling. Further, we have calculated the Fermi surface nesting function $\xi_{\mathbf q} = \sum_{\mathbf k} \delta(\epsilon_{\mathbf{k}}) \delta(\epsilon_{\mathbf{k+q}})$.~\citep{PhysRevB.82.184509,PhysRevB.88.064517} For all $\mathbf q$-vectors inside the Brillouin zone, the calculated $\xi_{\mathbf q}$ is substantially enhanced under the influence of Ag-tip [Figure \ref{f5}(b)]. Thus, increased electronic DOS, phonon softening along with increased phonon DOS, and substantial increase in $\xi_{\mathbf q}$ together contribute to increase in $\lambda$. Consequently, $\lambda(\omega \rightarrow \infty$) is increased by $\sim$ 67\% to 0.87 for the hcp-Zr/Ag-tip from 0.52 in pure hcp-Zr, which leads to substantial enhancement in the superconducting temperature (Table). 

In conclusion, we have shown that superconductivity of Zr gets dramatically enhanced under metallic point contacts. We observed a giant five-fold enhancement of superconducting critical temperature with an associated enhancement of superconducting energy gap. Our first principles calculations indicate that the enhancement is caused by enhanced density of states and altered electron-phonon coupling under the metallic point contacts.

GS acknowledges financial support from a grant awarded by SERB, DST under the grant number EMR/2015/001650. MA thanks CSIR for senior research fellowship (SRF). We acknowledge support from supercomputing facilities at the Centre for Development of
Advanced Computing, Pune; Inter University Accelerator Centre, Delhi; and at the Center for Computational Materials Science, Institute of Materials Research, Tohoku University. M. K. acknowledges funding from the Department of Science and Technology, Government of India under Ramanujan Fellowship, and Nano Mission
project SR/NM/TP-13/2016.
\\
\\



%

\end{document}